\documentclass[aps,prb,twocolumn, superscriptaddress, showpacs]{revtex4}

\usepackage{amsmath} \usepackage{amssymb} \usepackage{graphicx} \usepackage{pstricks}
\usepackage{scalefnt} \usepackage{color} \usepackage{ifpdf} \usepackage{pgffor}
\usepackage{verbatim} \usepackage{calc}  \usepackage{tikz}
\usepackage[all,arc,knot,rotate]{xy} \SelectTips{lu}{12} \usepackage{subfigure}

\newcommand{\ket}[1]{|{#1}\rangle} 
 \newcommand{\bpm}{\begin{pmatrix}}
\newcommand{\epm}{\end{pmatrix}} \newcommand{\bmm}{\begin{matrix}}
\newcommand{\emm}{\end{matrix}}

\newcommand{\TriVertex}{
  {
  \begin{tikzpicture}[scale=0.6]
  \tikzstyle{every node}=[font=\small]
  \draw  (0,0) -- (1,0.5);
  \draw  (2,0) -- (1,0.5);
  \draw  (1,1.5) -- (1,0.5);
  \node at(0.5,0.6) {$j_1$};
  \node at(1.7,0.6) {$j_2$};
  \node at(1.3,1.1) {$j_3$};
  \node at(1,0.15) {$v$};
  \end{tikzpicture}
  }}

\newcommand{\TrigPlaq}[3]{
  {
  \begin{tikzpicture}[scale=0.25]
  \tikzstyle{every node}=[font=\small]
  \draw  (0,-1.2) -- (0,0);
  \draw  (-2,3)  -- (0,0);
  \draw  (0,0)  -- (2,3);
  \draw  (2,3)  -- (-2,3);
  \draw  (-3,4) -- (-2,3);
  \draw  (3,4)  -- (2,3);
  \node at  (-1.7,1.3) {$#1$};
  \node at  (1.6,1.3) {$#2$};
  \node at  (0,3.7) {$#3$};
  \node at  (0.7,-0.6) {$j_4$};
  \node at  (-3,3.2) {$j_5$};
  \node at  (3,3.2) {$j_6$};
  \node at (0,1.6) {$p$};
  \end{tikzpicture}
  }}

\newcommand{\ManyFluxonBasisA}[8]{
 \bmm\xy
 0;/r0.5pc/:;
 (0,4)*{}="p1";
 (5,4)*\dir{*}="p2";
 (5,0)*{}="p3";
 (10,4)*\dir{*}="p4";
 (10,0)*{}="p5";
 (15,4)*{}="p6";
 (20,4)*{}="p7";
 (25,4)*\dir{*}="p8";
 (25,0)*{}="p9";
 (30,4)*{}="p10";
 "p1";"p2"**\dir{-}?(0.5)*\dir{>}+(0,1)*{\scriptstyle #1};
 "p3";"p2"**\dir{-}?(0.5)*\dir{>}+(1,-1)*{\scriptstyle #2};
 "p5";"p4"**\dir{-}?(0.5)*\dir{>}+(1,-1)*{\scriptstyle #3};
 "p9";"p8"**\dir{-}?(0.5)*\dir{>}+(1,-1)*{\scriptstyle #4};
 "p10";"p8"**\dir{-}?(0.5)*\dir{>}+(0,1)*{\scriptstyle #5};
 "p2";"p4"**\dir{-}?(0.5)*\dir{>}+(0,1)*{\scriptstyle #6};
 "p4";"p6"**\dir{-}?(0.5)*\dir{>}+(0,1)*{\scriptstyle #7};
 "p7";"p8"**\dir{-}?(0.5)*\dir{>}+(0,1)*{\scriptstyle #8};
 "p6";"p7"**\dir{--};
 \endxy\emm
 }

\newcommand{\ManyFluxonBasisB}[8]{
 \bmm\xy
 0;/r0.5pc/:;
 (0,0)*{}="p1";
 (5,0)*\dir{*}="p2";
 (5,4)*{}="p3";
 (10,0)*\dir{*}="p4";
 (10,4)*{}="p5";
 (15,0)*{}="p6";
 (20,0)*{}="p7";
 (25,0)*\dir{*}="p8";
 (25,4)*{}="p9";
 (30,0)*{}="p10";
 "p1";"p2"**\dir{-}?(0.5)*\dir{<}+(0,1)*{\scriptstyle #1};
 "p3";"p2"**\dir{-}?(0.5)*\dir{<}+(1,1)*{\scriptstyle #2};
 "p5";"p4"**\dir{-}?(0.5)*\dir{<}+(1,1)*{\scriptstyle #3};
 "p9";"p8"**\dir{-}?(0.5)*\dir{<}+(1,1)*{\scriptstyle #4};
 "p10";"p8"**\dir{-}?(0.5)*\dir{<}+(0,1)*{\scriptstyle #5};
 "p2";"p4"**\dir{-}?(0.5)*\dir{>}+(0,1)*{\scriptstyle #6};
 "p4";"p6"**\dir{-}?(0.5)*\dir{>}+(0,1)*{\scriptstyle #7};
 "p7";"p8"**\dir{-}?(0.5)*\dir{>}+(0,1)*{\scriptstyle #8};
 "p6";"p7"**\dir{--};
 \endxy\emm
 }

\newcommand{\TorusFluxonBasisA}[7]{ \xy
 0;/r0.5pc/:;
 (5,0)*\dir{*}="p2";
 (5,-4)*{}="p3";
 (10,0)*\dir{*}="p4";
 (10,-4)*{}="p5";
 (15,0)*{}="p6";
 (20,0)*{}="p7";
 (25,0)*\dir{*}="p8";
 (25,-4)*{}="p9";
 "p3";"p2"**\dir{-}?(0.5)*\dir{<}+(1,1)*{\scriptstyle #1};
 "p5";"p4"**\dir{-}?(0.5)*\dir{<}+(1,1)*{\scriptstyle #2};
 "p9";"p8"**\dir{-}?(0.5)*\dir{<}+(1,1)*{\scriptstyle #3};
 "p2";"p4"**\dir{-}?(0.5)*\dir{>}+(0,1)*{\scriptstyle #4};
 "p4";"p6"**\dir{-}?(0.5)*\dir{>}+(0,1)*{\scriptstyle #5};
 "p7";"p8"**\dir{-}?(0.5)*\dir{>}+(0,1)*{\scriptstyle #6};
 "p6";"p7"**\dir{--};
 "p2";"p8"**\crv{(-5,3)&(15,6)&(35,3)}?(0.5)*\dir{<}+(0,1)*{\scriptstyle #7};
 \endxy
 }

\newcommand{\TorusFluxonBasisB}[7]{ \xy
 0;/r0.5pc/:;
 (5,0)*\dir{*}="p2";
 (5,4)*{}="p3";
 (10,0)*\dir{*}="p4";
 (10,4)*{}="p5";
 (15,0)*{}="p6";
 (20,0)*{}="p7";
 (25,0)*\dir{*}="p8";
 (25,4)*{}="p9";
 "p3";"p2"**\dir{-}?(0.5)*\dir{<}+(1,1)*{\scriptstyle #1};
 "p5";"p4"**\dir{-}?(0.5)*\dir{<}+(1,1)*{\scriptstyle #2};
 "p9";"p8"**\dir{-}?(0.5)*\dir{<}+(1,1)*{\scriptstyle #3};
 "p2";"p4"**\dir{-}?(0.5)*\dir{>}+(0,1)*{\scriptstyle #4};
 "p4";"p6"**\dir{-}?(0.5)*\dir{>}+(0,1)*{\scriptstyle #5};
 "p7";"p8"**\dir{-}?(0.5)*\dir{>}+(0,1)*{\scriptstyle #6};
 "p6";"p7"**\dir{--};
 "p2";"p8"**\crv{(-5,-3)&(15,-6)&(35,-3)}?(0.5)*\dir{<}+(0,1)*{\scriptstyle #7};
 \endxy
 }

\begin{document}

\title{Emergent Exclusion Statistics of Fibonacci Anyons in 2D Topological Phases}

\author{Yuting Hu} \email{yuting@physics.utah.edu} \affiliation{Department of Physics and
Astronomy, University of Utah, Salt Lake City, UT 84112, USA}

\author{Spencer D. Stirling} \email{stirling@physics.utah.edu} \affiliation{Department of
Physics and Astronomy,  University of Utah, Salt Lake City, UT 84112, USA}
\affiliation{Department of Mathematics, University of Utah, Salt Lake City, UT 84112, USA}

\author{Yong-Shi Wu} \email{wu@physics.utah.edu} \affiliation{Key State Laboratory of Surface
Physics, Department of Physics \\ and Center for Field Theory and Particle Physics, Fudan
University, Shanghai 200433, China} \affiliation{Department of Physics and Astronomy,
University of Utah, Salt Lake City, UT 84112, USA}

\date{\today}

\begin{abstract} We demonstrate how the generalized Pauli exclusion principle emerges for
quasiparticle excitations in 2d topological phases. As an example, we examine the Levin-Wen
model with the Fibonacci data (specified in the text), and construct the number operator for
fluxons living on plaquettes. By numerically counting the many-body states with fluxon number
fixed, the matrix of exclusion statistics parameters is identified and is shown to depend on
the spatial topology (sphere or torus) of the system. Our work reveals the structure of the
(many-body) Hilbert space and some general features of thermodynamics for quasiparticle
excitations in topological matter.

\end{abstract}


\maketitle

\noindent{\bf 1. Introduction}

By now it is well-known that (quasi-)particles in strongly entangled many-body systems may
exhibit exotic quantum statistics, other than the familiar Bose-Einstein and Fermi-Dirac
ones. In addition to the anyonic or exchange statistics \cite{Wilczeck} in two dimensional
systems, statistical weight of many-body quantum states may also obey new combinatoric
counting rules \cite{Wu2} following a generalized Pauli exclusion principle \cite{Haldane},
in which the number of available single-particle states, when adding one more quasi-particle
into the system, linearly depends on the number of existing quasi-particles. A typical new
feature is mutual exclusion between different species, resulting in a matrix of statistical
parameters \cite{Haldane} and leading to unusual thermodynamics for ideal gases with only
statistical interactions\cite{Wu2,footnote1}. (For a review see, e.g., ref.
\onlinecite{Wu1994}.)

More precisely, following ref. \onlinecite{Wu2}, in the case with only one species of
quasi-particles, the number of $N$-particle states is assumed to be given by the binomial
coefficient: \begin{equation}
  \label{statecounting}
  W_{G,N}=\binom{G_{\text{eff}}+(N-1)}{N},
\end{equation} with $G_{\text{eff}}=G-\alpha(N-1)$
being the number of available single-particle states, while $G$ is the number of
single-particle states when $N=1$. Then $\alpha=0$ corresponds to bosons and $\alpha=1$
fermions; other values of $\alpha$ gives rise to exotic exclusion statistics. Similarly, in
the multi-species case, the number of many-particle states is assumed to be given by
($a,b=1,\dots,m$ labeling species) \begin{equation}
  \label{multispecies}
  W_{\{G_a,N_a\}}=\prod_{a}
  \binom{G_a+N_a-1 -\sum _{b=1}^{m} \alpha _{ab}
  \left(N_b-\delta_{ab}\right)}{N_a}.
\end{equation} Here coefficients $\alpha_{ab}$ form the (mutual) statistics matrix.

It has been shown \cite{WuBernard} that the Thermodynamic Ansatz \cite{YangYang} for
one-dimensional solvable many-particle models is actually a special case of the exotic
exclusion statistics. (See also refs. \onlinecite{Ha,Kohmoto}.) It has been also numerically
verified that quasi-particle excitations in the fractional quantum Hall (FQH) systems indeed
obey \cite{Zhang} eq. (\ref{statecounting}), or eq. (\ref{multispecies}) allowing mutual
exclusion between different species\cite{SuWuYang}. Moreover either the Haldane or Jain
hierarchy in the FQH effect can be theoretically understood from the exclusion statistics of
quasiparticles\cite{Wu1994,YuYue}.

Recently there has been revived interest in the study of quasiparticle statistics in 2d
topological states of matter (including FQH systems), because of the possibility of using
their braiding to do (fault tolerant) topological quantum computation (TQC)
\cite{Kitaev,Wang}. In order to know better about the error of TQC at finite temperature, it
is needed to understand better how exclusion statistics of quasi-particles emerges in 2d
topological matter, which governs the thermodynamics of the system.

In this letter, we carry out the many-body state counting in an exactly solvable discrete
model, i.e., the Levin-Wen model\cite{LW} (with a special set of data), that describes a 2d
topological quantum fluid\cite{Gil0906} of Fibonacci anyons\cite{SB}, with doubled Fibonacci
anyons as fluxon excitations living on plaquettes. The Fibonacci anyons are the simplest
non-abelian anyons. They occur as quasiparticles in the $k=3$ Read-Rezayi state\cite{RR} in
an FQH state with filling fraction $\nu=\frac{12}{5}$, and can be used for universal
topological quantum computation\cite{Wang}. (Recently, it is proposed\cite{LK} that the
physics of interacting Fibonacci anyons may be studied in a Rydberg lattice gas.)

In this Letter, we first construct the number operator for fluxons in the model, which helps
us identify the states with localized excitations. Then we numerically count the (many-body)
states with fluxon-number $N$ fixed, from $N=1$ up to $N=7$, for the system on a sphere and
torus, respectively. The results exhibit a pattern closely related to the Fibonacci numbers,
which in turn is put in the form of eq. (\ref{multispecies}), thus determining a
topology-dependent statistics parameter matrix. Our work reveals that exotic exclusion
emerges among quasiparticles due to interplay between various ``hidden'' degrees of freedom
(d.o.f.) in addition to fluxon locations. These ``hidden'' d.o.f. are very similar to the
pseudo-species, previously introduced in the literature on conformal field theory\cite{GS},
which do not contribute to energy but contribute to state-counting in accordance with an
exclusion statistics parameter matrix. Finally, we briefly discuss the thermodynamics of the
system.

\smallskip

\noindent{\bf 2. The model}

We consider a discrete model for a ``spin'' system on a trivalent graph on a closed surface,
e.g. a sphere or torus. We adopt a simplified formulation of the Levin-Wen model \cite{LW},
with the Fibonacci data (e.g., see ref. \onlinecite{Wang}) given as follows: Each link is
assigned a ``spin''-type labeled by $j$ and configurations of the labels on all links form an
orthonormal basis in the Hilbert space. The key input of the Fibonacci data is that the
``spin"-type index $j$ takes only two values $j=0,1$, and they satisfy an algebra (called the
Fibonacci algebra), which describes how to fuse two ``spin"-types through the branching
rules: \begin{equation} \label{fusionrules} 0\otimes j=j\otimes 0=j, 1\otimes 1= 0\oplus 1.
\end{equation} These rules are similar to those for the (direct sum) decomposition of
(tensor) products of irreducible representations of a group, with $j=0$ playing the role of
the unit element for the (tensor) product. (It is conjectured that the Levin-Wen model
describes a class of doubled (time-reversal invariant) topological phases \cite{Freedman}.)

The Hamiltonian of the model is of the form 
\begin{equation}
  \label{HamiltonianBps}
  \hat{H}=U\sum_{v}(1-\hat{Q}_v)+\epsilon\sum_{p}(1-\hat{B}_p),
  \quad
  \hat{B}_p=\frac{1}{D}\sum_{s=0,1}d_s\hat{B}_p^{s}.
\end{equation} 
The two summations here run over all vertices $v$ and plaquettes $p$,
respectively. For $\hat{B}_p$, the summation runs over the ``spin"-type $s=0,1$, and
$d_0=1$,$d_1=\phi\equiv{(\sqrt{5}+1)/2}$, and $D=1+\phi^2$. $U$ and $\epsilon$ are positive
constants. The explicit form of the operators $\hat{Q}_v$ and $\hat{B}_p^{s}$ are given in
the supplemental material\cite{suppl}. (By adding more competing interactions, ref.
\cite{Vidal} has used this model to discuss topological phase transitions in the Fibonacci
anyon liquid. Here we restrict to the original Levin-Wen model and discuss emergent exclusion
statistics for quasi-excitations.)

A notable property of the model is that by construction, $\hat{Q}_v$ and $\hat{B}_p$ are
mutually commuting projection operators: $[ \hat{Q}_v,\hat{B}_{p}]=0$,
$\hat{Q}_v\hat{Q}_{v'}=\delta_{vv'}\hat{Q}_v$ and
$\hat{B}_p\hat{B}_{p'}=\delta_{pp'}\hat{B}_p$. Thus the Hamiltonian is exactly solvable. The
energy eigenstates are the simultaneous eigenvectors of these projections $\hat{Q}_v$ and
$\hat{B}_p$. The ground states are those $\ket{\Phi}$ that satisfy
$\hat{Q}_v\ket{\Phi}=\ket{\Phi}=\hat{B}_p\ket{\Phi}$, for all $v$ and $p$. Using the method
developed in ref. \onlinecite{GSD}, one can compute ground state degeneracy: $\text{GSD}=1$
on sphere and $\text{GSD}=4$ on torus.

\begin{figure}\centering
  \includegraphics{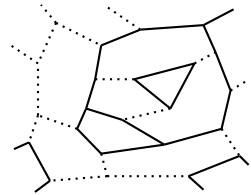}
  \caption{A configuration in the subspace with only fluxons allowed, with
  solid lines for $j=1$ and dotted lines $j=0$. Open strings (with only
  single solid line at some vertex) are forbidden by $\delta_{001}=0$.}
  \label{fig1}
\end{figure}

The quasiparticle excitations are the states with zero eigenvalue of $\hat{Q}_{v'}$ for some
$v'$ and/or of $\hat{B}_{p'}$ for some $p'$. 
$\hat{B}_{p'}\ket{\Psi}=0$). In this letter we restrict ourselves to study the so-called
fluxons, satisfying $\hat{Q}_{v}=1$ for all $v$, $\hat{B}_{p'}=0$ for a specified set of $p'$
(where fluxons live), and $\hat{B}_{p}=1$ for all other $p$. (See Fig. \ref{fig1}.)

\smallskip

\noindent{\bf 3. Number operator of fluxons}

A crucial property of the model is that ${\hat{B}_p^s}$ defined above form an abelian
algebra\cite{LW}: \begin{equation}
  \label{BpsAlgebra}
  \hat{B}_p^r\hat{B}_{p}^s=\sum_{t=0,1}\delta_{rst}\hat{B}_p^t,
\end{equation} Here $\delta_{ijk}=\delta_{jki}=\delta_{jik}$ is given by
$\delta_{000}=\delta_{011}=\delta_{111}=1$, arising from the branching rules
\eqref{fusionrules}. From the generators we construct the operators ($i=0,1$):
\begin{equation}
  \label{topologicalcharge}
  \hat{n}_p^i=\sum_{j=0,1}S_{i0}S_{ij}\hat{B}_p^j,
\end{equation} where $S$ is the modular matrix given by \begin{equation}
  \label{Smatrix}
  S=\frac{1}{\sqrt{D}}
    \left(
    \begin{array}{cc}
     1 & \phi  \\
     \phi  & -1
    \end{array}
    \right),
\end{equation} with $S_{ij}$ for fixed $i$ being a one-dimensional representation of the
algebra \eqref{BpsAlgebra}. One can check that $\hat{n}_p^i$ ($i=0,1$) form a complete set of
orthonormal projections: \begin{equation} \label{orthoN}
\hat{n}_p^i\hat{n}_p^k=\delta_{ik}\hat{n}_p^k, \quad \hat{n}_p^0+\hat{n}_p^1=\mathbf{1}.
\end{equation}

There is a fluxon at $p$ in a state $\ket{\Psi}$, if $\hat{n}_p^1\ket{\Psi}=\ket{\Psi}$. A
ground state $\ket{\Phi}$ contains no fluxon because $\hat{n}_p^0=\hat{B}_p$. Hence the model
has only one type of fluxons, and there is no state with two fluxons living at the same
plaquette $p$. This seems to indicate that the flux in this model should be a fermion (or a
hard boson). In the following we will present a study of many-fluxon state counting in this
model, which reveals that actually the fluxons in this model obey instead exotic exclusion
statistics which was proposed in refs. \onlinecite{Haldane,Wu2}. (See also the
footnote\cite{footnote}.)

\smallskip

\noindent {\bf 4. Exclusion statistics on a sphere}

Let us count the $N$-fluxon states in the model with $P$ plaquettes on a sphere. Pick up a
set of $N$ fixed plaquettes and denote it by $\mathcal{C}=\{p_1,p_2,...,p_N\}$ ($N<P$). The
states with exactly $N$ fluxons occupying the selected plaquettes are those $\ket{\psi}$
satisfying \begin{align}
  \label{Nstate}
  &\hat{n}_p^j\ket{\Psi}=\delta_{j1}\ket{\Psi},
  \quad \text{ for }p\in{\mathcal{C}},
  \nonumber\\
  &\hat{n}_{p'}^j\ket{\Psi}=\delta_{j0}\ket{\Psi},
  \quad \text{ for }{p'}\notin{\mathcal{C}}.
\end{align}

Thus $\left(\prod_{p\in{\mathcal{C}}}\hat{n}_p^1
\prod_{p'\notin{\mathcal{C}}}\hat{n}_{p'}^0\right)$ is the projector onto the subspace of
such states. Tracing this projection computes the total number of the $N$-fluxon states in
the configuration $\mathcal{C}$: \begin{equation}
  \label{counttrace}
  w_{P,N,\mathcal{C}}=\text{tr}(\prod_{p\in{\mathcal{C}}}
\hat{n}_p^1\prod_{p'\notin{\mathcal{C}}}\hat{n}_{p'}^0). \end{equation}

We numerically compute eq. \eqref{counttrace} on random graphs on sphere with $P(\geq7)$
plaquettes, with the stable result presented in Table \ref{table:sphereCounting}.
\begin{table}[ht] \caption{State Counting on Sphere}\label{table:sphereCounting} \centering
\begin{tabular}{lcrrrrrrrr} \hline\hline Fluxon number $N$ & & 0 & 1 &  2 &  3 &  4 &  5  & 6
& 7 \\ \hline State Counting $w_{P,N,\mathcal{C}}$ & & 1 & 0 & 1 & 1 & 4 & 9 & 25 & 64\\
\hline \end{tabular} \end{table}

The pattern of the $N$-dependence is obvious: \begin{equation}
  \label{wNC}
  w_{P,N,\mathcal{C}}=F_{N-1}^2,
\end{equation} where $F_n$ is the Fibonacci number that satisfies the recurrence relation
$F_n=F_{n-1}+F_{n-2}$ with $F_1=F_2=1$. Both numerically and analytically we have checked
that eq. \eqref{wNC} is independent of the graph, of the total number $P$ of plaquettes, as
well as the locations of the $N$ fluxons. The appearance of the squared in eq. \eqref{wNC} is
consistent with the conjecture that the LW model describes a {\em doubled} topological phases
\cite{Freedman,GSD}.

Summing over configurations $\mathcal{C}$ (i.e., over possible distributions of N plaquettes
in a fixed graph), we get the total number of $N$-fluxon states: \begin{equation}
  \label{counting}
  W_{P,N}^{\text{sphere}}
=\sum_{\mathcal{C}}w_{P,N,\mathcal{C}} =\binom{P}{N}F_{N-1}^2. \end{equation}

The first factor counts the ways to distribute $N$ fluxons over $P$ plaquettes. The second
factor counts the states of the link d.o.f., which are not unique, given $N$ and
$\mathcal{C}$. The independence of $w_{P,N,\mathcal{C}}$ on $P$ and $\mathcal{C}$ implies the
degeneracy of the excited states is topological in the sense that it does not depend on the
detailed structure of the underlying graph, and not on the relative positions between the
fluxons as well. We have numerically check particularly this property (see the supplemental
meterial \cite{suppl}£©. The origin of this property lies in the topological symmetry of the
model under mutations of the underlying graph\cite{GSD}.

To find the exclusion statistics, we rewrite \eqref{counting}: \begin{equation}
  \label{binomcounting}
  W_{P,N}^{\text{sphere}}=\binom{P}{N}
  \sum_{N_1,N_2=0}^{[\frac{1}{2}(N-2)]}
  \binom{N-N_1-2}{N_1} \binom{N-N_2-2}{N_2},
\end{equation} where $[x]$ is the greatest integer less than or equal to $x$.

Now eq. \eqref{binomcounting} is of the form of eq. \eqref{multispecies}, by introducing two
additional pseudo-species $a=1,2$, which do not contribute to the total energy but are
helpful for state-counting. This is similar to what was suggested for state counting in some
conformal field theories \cite{GS}. Including the original fluxon species labeled by $a=0$,
from eq. \eqref{binomcounting} we read the exclusion statistics parameters $\alpha_{ab}$
($a,b,=0,1,2$): \begin{align}
  \label{spherestatistics}
  \alpha^{\text{sphere}}=\left(
    \begin{array}{ccc}
     1 & 0 & 0 \\
     -1 & 2 & 0 \\
     -1 & 0 & 2
    \end{array}
\right). \end{align}

The diagonal $\alpha_{aa}$ is the self-exclusion statistics for species $a$. The
$\alpha_{00}=1$ implies the hard-core boson behavior, that takes care of the first
combinatoric factor in $\binom{P}{N}$ in eq. \eqref{counting} and eq. \eqref{binomcounting}.
This can be understood with eq. \eqref{orthoN}.

The pseudo-species provides a way to count states, in the presence of fluxons, of link
d.o.f., which are not uniquely determined by the constraints \eqref{Nstate}. The value
$\alpha_{11}=\alpha_{22}=2$ implies that one pseudo-particle makes two single-particle states
(or ``seats'') unavailable to an additional pseudo-particle. The negative mutual statistics
$\alpha_{20}=\alpha_{30}=-1$ tells us that each fluxon present creates one vacant ``seat''
for each pseudo-species. So the maximum particle number of each pseudo-species is naturally
$[(N-1)/2]$. These results help us understand the structure of the (many-body) Hilbert space
for excited states of the system, and help derive analytically the state counting formula
\eqref{binomcounting}. (A sketch of such a derivation is presented in the supplemental
material\cite{suppl}.)

We note that the many-body counting formula eq. \eqref{multispecies}, proposed in ref.
\onlinecite{Wu2}, with the statistical matrix \eqref{spherestatistics}, exactly reproduces
the result, Table \ref{table:sphereCounting}, of numerical counting for fluxon numbers from
$N=0$ to $N=7$. It is remarkable that the counting formula eq. \eqref{multispecies} is valid
even for very small values of the fluxon number, so we believe it is an exact result, true
for all values of $N$, including the thermodynamical limit.

\smallskip

\noindent {\bf 6. Exclusion statistics on a torus}

We proceed to consider the model on a torus. The ground state degeneracy\cite{GSD} is 4 .
Thus the system exhibits the global topological d.o.f., and we can study their effects on
excited states by counting the pseudo-particle states.

Pick up $N$ plaquettes ($N<P$). The number of states with $N$ fluxons on htese plaquettes is
computed numerically as in Table \ref{table:torusCounting}. \begin{table}[ht] \caption{State
Counting on Torus} \label{table:torusCounting} \centering \begin{tabular}{lclllllll}
\hline\hline Fluxon number $N$ & & 0 & 1 &  2 &  3 &  4 &  5  & 6 \\ \hline State Counting  &
& $2^2$ & 1 & $3^2$ & $4^2$ & $7^2$ & $11^2$ & $18^2$\\ \hline \end{tabular} \end{table}

The pattern of its dependence on $N$ is \begin{equation}
  \label{Lucascounting}
  W_{P,N}^{\text{torus}}=\binom{P}{N}L_{N}^2,
\end{equation} with $L_n$ the Lucas number, a modified version of the Fibonacci number,
satisfying the recurrence relation $L_n=L_{n-1}+L_{n-2}$ with $L_1=1, L_2=3$.

We rewrite \eqref{Lucascounting} in terms of binomial coefficients: \begin{align}
  \label{torusbinomcounting}
  W_{P,N}^{\text{torus}}=
  &\binom{P}{N}\sum_{N_1,N_2=0,1}
  \binom{1}{N_1}\binom{1}{N_2}\times
  \nonumber\\
  &\sum_{N_3,N_4=0}^{[ \frac{1}{2}(N-2)]}
  \binom{N-2N_1-N_3}{N_3} \binom{N-2N_2-N_4}{N_4},
\end{align} and get the exclusion statistics parameters $\alpha_{ab}$ ($a,b=0,1,2,3,4$):
\begin{align}
  \label{torusstatistics}
  \alpha^{\text{torus}} =\left(
    \begin{array}{ccccc}
     1 & 0 & 0 & 0 & 0 \\
     0 & 1 & 0 & 0 & 0 \\
     0 & 0 & 1 & 0 & 0 \\
    -1 & 2 & 0 & 2 & 0 \\
    -1 & 0 & 2 & 0 & 2
    \end{array}
    \right),
\end{align} where we denote by $a=0$ the fluxon species.

Eq. \eqref{torusbinomcounting} shows that one needs to introduce four pseudo-species
$a=1,2,3,4$. The pseudo-species $a=1,2$ are interpreted as the topological d.o.f. on the
torus, for the following reasons. The allowed ``particle number'' $N_1,N_2=0,1$ of these
pseudo-species are independent of the number $N$ of fluxons. Particularly when there is no
fluxon present, the configurations $N_1,N_2=0,1$ characterize the four-degenerate ground
states. Then the pseudo-species $a=3,4$ provide a way to count the states of link d.o.f.
given a ground state and fluxon number.

The state counting of excitations on a torus is shown different from that on a sphere. (A
state counting formula of different form from ours, which also exhibits the dependence on the
spatial topology, is reported in ref. \onlinecite{Vidal}, without making connection to
exclusion statistics.) Indeed the mutual statistics parameters $\alpha_{31}=\alpha_{42}=2$
imply that the number of states of link d.o.f. $a=3$ ($a=4$) are affected by the topological
d.o.f. $a=1$ ($a=2$), respectively. On the other hand, the topological d.o.f. are not
affected by the fluxons present and the link d.o.f.. So the degenerate ground states can be
used to label the sectors of excitations. We note that in the sector with $N_1=N_2=1$, the
state counting for fluxons is exactly the same as that on sphere.

\smallskip

\noindent {\bf 7. Statistical Thermodynamics}

Now we assume that only fluxons can be thermally excited; this is the case when $U>>\epsilon,
kT$ in eq. \eqref{HamiltonianBps}. In the thermodynamic limit, the Hilbert space dimension of
$N$-fluxon states (occupying $N$ fixed plaquettes) is asymptotically \begin{align}
  \label{TheormoLimitDimH}
  &\text{on sphere: }
  \quad\lim_{N\to \infty}\,F_{N-1}^2\sim \phi^{2N-2}/5,
  \nonumber\\
  &\text{on torus: }
 \quad\quad\lim_{N\to \infty}\,L_N^2\sim\phi^{2N}.
\end{align} ($\phi^2$ is called the quantum dimension of the fluxon.) On a torus, for
example, the canonical partition function is \begin{equation}
  \label{partitionfunction}
  Z^{\text{torus}}=\sum_{N=0}^{P}\binom{P}{N}L_N^2e^{-N\epsilon/kT}
  \sim(\phi^2e^{-\epsilon/kT}+1)^P.
\end{equation}

It can be interpreted as the grand canonical partition function of the many-fluxon system,
which behaves like a fermionic system with a {\em temperature-independent fugacity} $z$ given
by the quantum dimension: \begin{equation} z=\phi^2. \end{equation}

The fugacity $z$ counts the effective number of states per fluxon located at a plaquette.
Note that $z$ is irrational rather than integer. This is a manifestation that the many-fluxon
states are highly entangled ones with long-range entanglement. They are superpositions of
highly constrained $j$-configurations on the links, obviously not of the form of a direct
product of localized fluxon states.

product 
thermodynamic behavior 
example, 
energy of any $N$-fluxon state is equal to 
with the chemical potential set to zero.

The statistical distribution of the average occupation number of fluxons is obtained from eq.
\eqref{partitionfunction}: \begin{equation}
  \label{distr}
  \langle n\rangle=\langle N\rangle/P
  =\frac{1}{e^{\epsilon/kT}{\phi^{-2}}+1}.
\end{equation} Many useful thermodynamic observables are then computable. The probability for
thermal excitations of fluxons that cause errors in topological quantum computation which
uses the code based on this model can then be estimated more accurately than before.

Though the model is very simple, we believe that the features revealed in this letter should
be quite general for emergent exotic exclusion statistics and thermodynamics for
quasiparticle excitations in a wide class of 2d topological phases. Moreover, the knowledge
and insights gained in this model for the Hilbert space structure of many-fluxon states may
be useful in the future for fault-tolerant quantum computation codes and alogrithms that
explore systems in topological phases.

\smallskip

\noindent {\bf Acknowledgement:} YH thanks Department of Physics, Fudan University for warm
hospitality he received during a visit in summer 2011 and 2012. YSW was supported in part by
US NSF through grant No. PHY-1068558.

\onecolumngrid

\clearpage

\begin{center} {\bf \large Supplementary Material} \end{center}

\section{Explicit Hamiltonian of the Levin-Wen Model}

The operator $\hat{Q}_v$ defined at vertex $v$ in the Hamiltonian \eqref{HamiltonianBps} in
the text is 
\begin{align}
  \hat{Q}_v\left|\bmm
  \TriVertex
  \emm\right\rangle
  =\delta_{j_1j_2j_3}\left|\bmm
  \TriVertex
  \emm\right\rangle.
\end{align} 
(Only the relevant part of the graph is shown; the rest of the graph is the same
on both sides.) Here $\delta_{ijk}=\delta_{jki}=\delta_{jik}$ is given by 
\begin{equation}
  \label{branchingrule}
  \delta_{000}=\delta_{011}=\delta_{111}=1,\delta_{001}=0
\end{equation} 
(called the Fibonacci fusion rule\cite{Wang}).

The operator $\hat{B}_p^s$ defined at plaquette $p$ in the Hamiltonian \eqref{HamiltonianBps}
in the text is 
\begin{align} \label{Bps} &\Biggl\langle\bmm \TrigPlaq{j'_1}{j'_2}{j'_3}
\emm\Biggr| \hat{B}_p^s \Biggl|\bmm \TrigPlaq{j_1}{j_2}{j_3} \emm\Biggr\rangle\nonumber\\ =&
v_{j_1}v_{j_2}v_{j_3}v_{j'_1}v_{j'_2}v_{j'_3}
G^{j_5j_1j_3}_{sj'_3j'_1}G^{j_4j_2j_1}_{sj'_1j'_2} G^{j_6j_3j_2}_{sj'_2j'_3}, \end{align}
where $v_j\equiv{\sqrt{d_j}}$. A similar rule applies when the plaquette $p$ is a quadrangle,
a pentagon, or a hexagon etc. Note that the matrix is nondiagonal only on the labels of the
boundary links (i.e., $j_1$, $j_2$, and $j_3$ on the above graph).

In the Fibonacci data, the non-vanishing $6j$-symbol $G$'s are given by 
\begin{align}
  \label{6js}
  G^{000}_{000}=1,
  G^{011}_{011}=G^{011}_{111}=1/\phi,
  \nonumber\\
  G^{000}_{111}=1/\sqrt{\phi},
  G^{111}_{111}=-1/{{\phi}^2},
\end{align} 
with the (tetrahedral) symmetry: 
\begin{align} \label{6jsymm}
&G^{ijm}_{kln}=G^{mij}_{nkl} =G^{klm}_{ijn}=G^{jim}_{lkn}. \end{align} 
One can check that
they satisfy the conditions: 
\begin{align} 
\label{6jcond} 
&\text{pentagon id:}
\quad\quad\sum_{n=0,1}{d_{n}}G^{mlq}_{kpn}G^{jip}_{mns}G^{jsn}_{lkr}
=G^{jip}_{qkr}G^{riq}_{mls},\nonumber\\ &\text{orthogonality:}
\quad\sum_{n=0,1}{d_{n}}G^{mlq}_{kpn}G^{lmi}_{pkn}
=\frac{\delta_{iq}}{d_{i}}\delta_{mlq}\delta_{kip}. 
\end{align}

These expressions and properties can be used to prove that $\hat{Q}_v$ and $\hat{B}_p$ are
mutually commuting projection operators. Thus the Hamiltonian \eqref{HamiltonianBps} is
exactly solvable. (See the text.) \smallskip

\section{Numerical verification}

The state counting is numerically computed by exactly diagonalization of the Hamiltonian
\eqref{HamiltonianBps} in the $Q_v=1$ subspace: the number of $N$-fluxon states is the
number of $E=N\epsilon$ eigenvalues. We choose random graphs, with the total number of
plaquettes up to $P=7$ on a sphere and up to $P=6$ on a torus.

To verify $w_{P,N,\mathcal{C}}=F_{N-1}^2$ on a sphere (and $L_N^2$ on a torus), we
numerically check the following topological properties: (1) if we fix the graph, we can count
the number of states with $N$ fluxons at $N$ fixed plaquettes (with $n^{j=1}_p=1$ at these fixed
plaquettes and $n^{j=1}_p=0$ at the rest ones), which does not depend on where the $N$ fixed
plaquettes are chosen; (2) if we choose different graphs with the fixed total number $P$ of plaquettes, the number of states does not
depend which graph we choose; (3) and if we choose different graphs with different total number $P$ of plaquettes, the number of states does not depend on $P$
(as long as $P\geq N$ with $N$ fixed). As a result, the state counting only depends the number $N$ of
fluxons.

\section{The analytic state counting}

\begin{figure}[h!]
  \centering
  \subfigure[]{\includegraphics[scale=0.5]{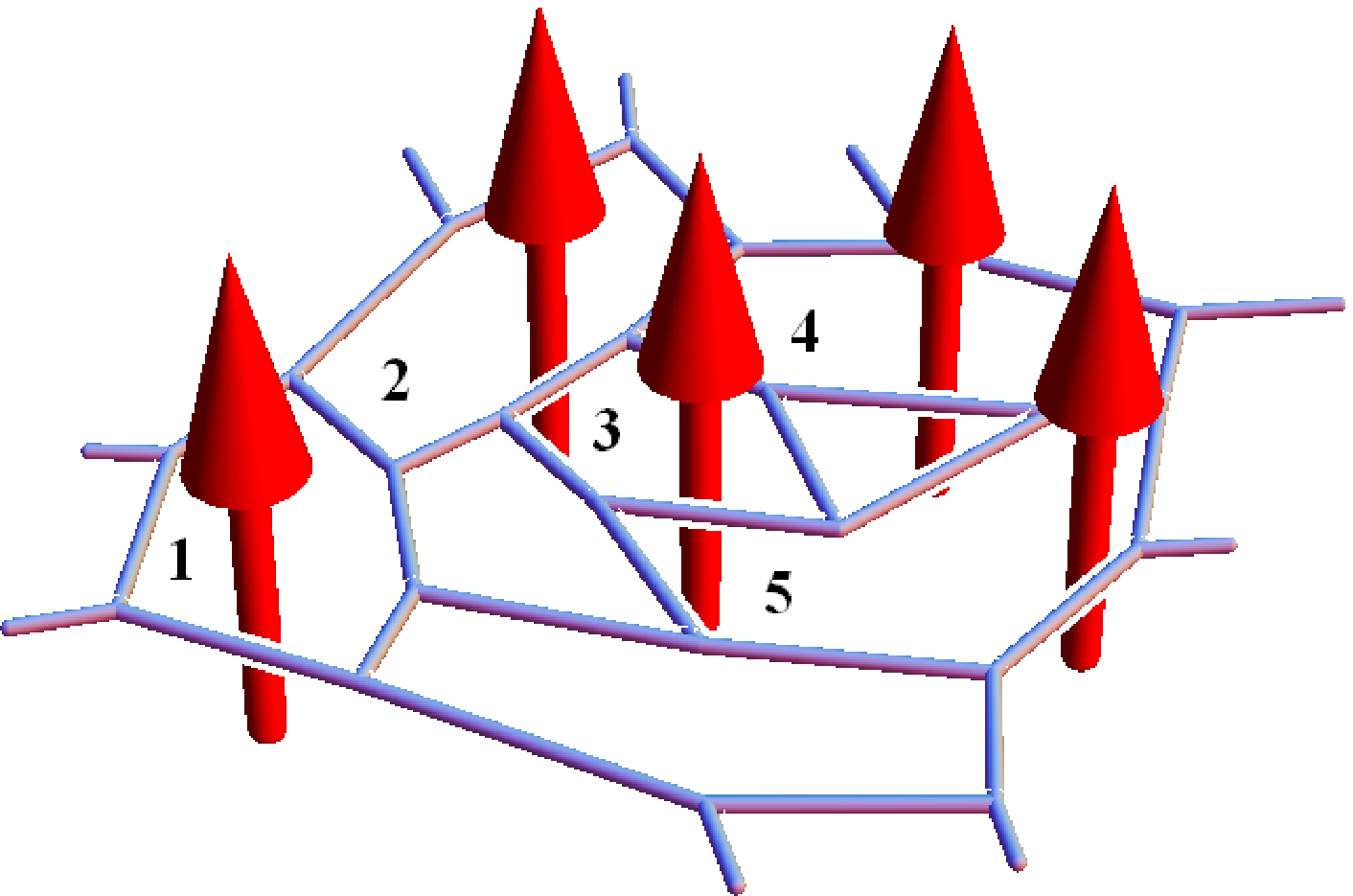}}
  \subfigure[]{\includegraphics[scale=0.5]{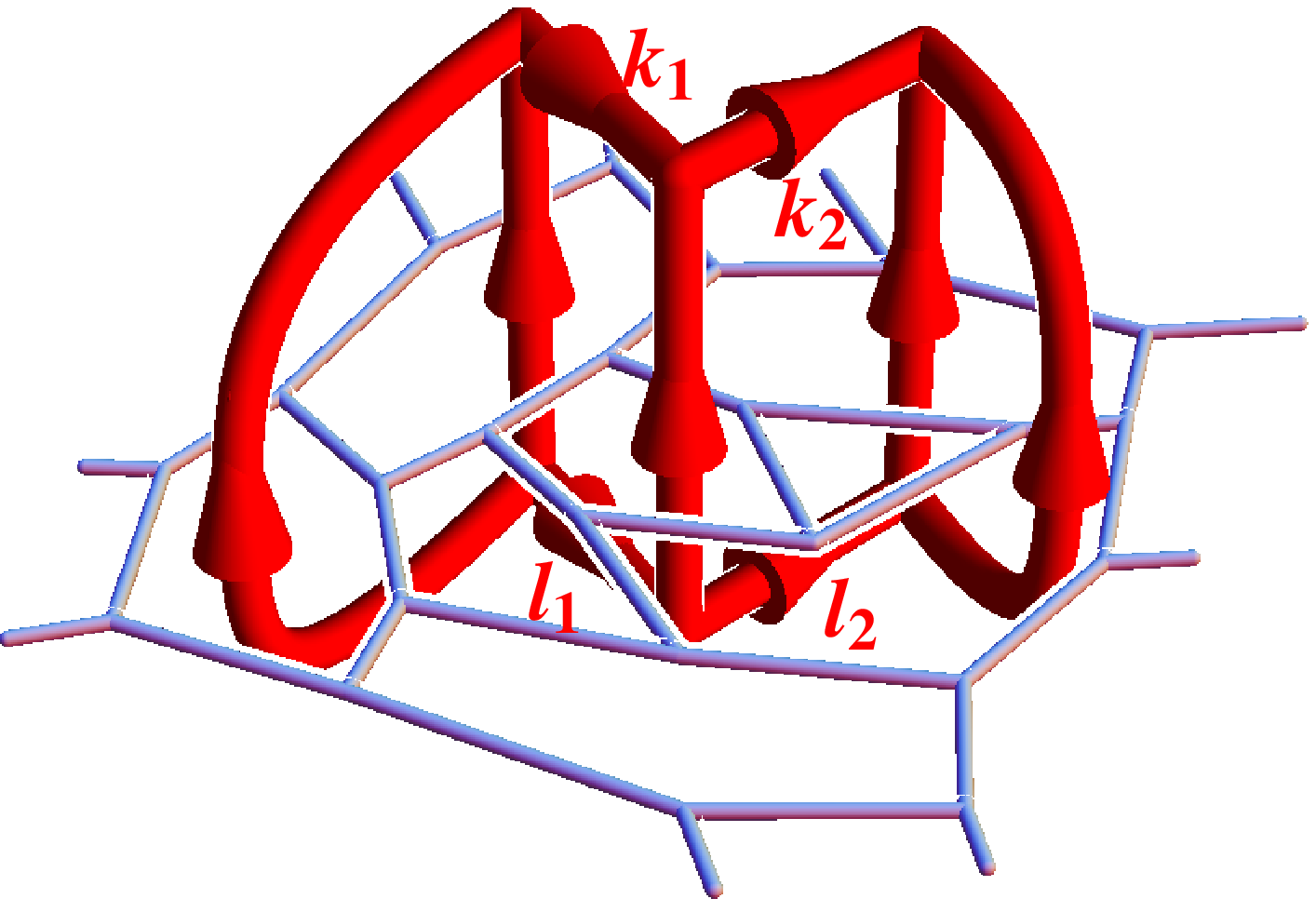}}
  \caption{(color online) A flux-tube network represents quantum numbers of an elementary
  excitation with $N=5$ fluxons. The trivalent graph is the same as the one in Figure
  \ref{fig1}, and is assumed to be on a sphere.}
  \label{fig2}
\end{figure}

Here we sketch how one can count states with $N$-fluxon excitations analytically, with
results in agreement with the numerical results reported above. The details will be presented
in Ref. \onlinecite{Hu2}, in which an operator approach will be developed to fully
characterize the quantum numbers of fluxon excitations. Here we just briefly present the
resulting description of the full set of quantum numbers for an elementary fluxon excitation
in terms of a flux-tube network; see, e.g., Fig. \ref{fig2}. In this figure we consider an
excitation state, say, with $N=5$, i.e. exactly five fluxons at the fixed plaquettes
$p=1,2,\dots,5$ on a sphere. Such a state carries eigenvalues $n^{j=1}_p=1$; namely a flux
labeled by $j=1$ pierces through each of the five plaquettes (see the five flux tubes in Fig.
\ref{fig2}(a)).

The above numerical exact diagonalization shows that with the five occupied plaquettes fixed,
there are more than one five-fluxon excitations. Therefore, the $n^j_p$'s are not enough to
give the full set of quantum numbers for the degenerate five-fluxon states. Moreover, the
numerical results also shows that the degeneracy does not depend the locations of the five
occupied plaquettes. Thus the degeneracy is topological in nature, and this suggests that the
quantum numbers, other than the $n^j_p$'s, that distinguish the degenerate five-fluxon states
should be also topological in nature. Physically these extra quantum numbers, which we call
as topological charges, describe relative degrees of freedom among the fluxons.

In Ref. \onlinecite{Hu2}, we will show that the topological charges for five-fluxon states
can be obtained by starting from the consideration of topological charges for a subsystem
consisting of only two fluxons. It turns out that the topological charges of two fluxons can
be classified by the quantum double charges of the Fibonacci data $\{j=0,j=1\}$. Following
Ref. \cite{Wang}), we denote the string types $j=0$ and $j=1$ by $\mathbf{1}$ and $\tau$,
respectively. Then the quantum double charges for a two-fluxon states are then denoted by
$(\mathbf{1},\overline{\mathbf{1}})$, $(\tau,\overline{\mathbf{1}})$,
$(\mathbf{1},\overline{\tau})$, and $(\tau,\overline{\tau})$.

The full set of quantum numbers for five-fluxon states in terms of quantum double charges can
be represented by flux-tube networks as those shown in Fig. \ref{fig2}. In addition to $n^{\tau}_p=1$ for $p=1,2,\dots,5$,
we have two more quantum numbers: the total topological charge $(k_1\overline{l_1})$ of the
subsystem containing two fluxons at $p=1$ and $2$, and $(k_2\overline{l_2})$ of the subsystem
containing three fluxons at $p=1,2$ and $3$. (These quantum numbers are defined in
\cite{Hu2}, which we will not dwell on in this letter). Both of them take values in the
quantum double charges, and can be thought as the flux connecting the $\tau$ flux through the
five fixed plaquettes. In Fig. \ref{fig2}, we have five $\tau$ fluxes through the five fixed
plaquettes, and the two $\tau$ fluxes through $p=1$ and $2$ couple to $k_1$ flux above the
plane and to $l_1$ below the plane. The allowed values of $k_1$ and $l_1$ are constrained by
the fusion rule $\delta_{\tau\tau k_1}=1$ and $\delta_{\tau\tau l_1}=1$, i.e.,
$k_1,l_1=\mathbf{1},\tau$. The fluxes $k_2$ and $l_2$ result from coupling $k_1$ and $l_1$ to
the $\tau$ flux through $p=3$.

This set of quantum numbers classify the degenerate five-fluxon states on a sphere, giving a
basis \begin{equation} \left\{\ket{k_1\overline{l_1},k_2\overline{l_2}}\right\},
\qquad\text{constrained by }\quad
\delta_{\tau k_1k_2}=1,\delta_{\tau l_1 l_2}=1.
\end{equation} 
$k_1,k_2$ (or
$l_1,l_2$) take three possible values: $\ket{k_1=\mathbf{1},k_2=\tau}$,
$\ket{k_1=\tau,k_2=\mathbf{1}}$, and $\ket{k_1=\tau,k_2=\tau}$. Therefore five-fluxon
excitations (at five fixed plaquettes) have the degeneracy $(F_{5-1})^2=9$.

In general, $N$-fluxon excitations on a sphere have a basis labeled by $N-3$ quantum-double
charges $\ket{k_1\overline{l_1},k_2\overline{l_2},\dots,k_{N-3}\overline{l_{N-3}}}$,
constrained by $\delta_{\tau k_n k_{n+1}}=1$ and $\delta_{\tau l_n l_{n+1}}=1$. To view them
clearly, we take the skeleton of the flux-tube network in Fig. \ref{fig2} as the basis:
\begin{align} \label{eq:basis}
&\ManyFluxonBasisA{\tau}{\tau}{\tau}{\tau}{\tau}{k_1}{k_2}{k_{N-3}} \quad\otimes\quad
\ManyFluxonBasisB{\tau}{\tau}{\tau}{\tau}{\tau}{l_1}{l_2}{l_{N-3}}, \end{align} where the
left part corresponds to the flux-tube network above the plane, and the right part to the one
below the plane. The external links correspond to the flux tubes through the plaquettes. The
basis depends on the ordering of the $N$ fluxons. In Fig. \ref{fig2}, if we choose different ordering of
fluxons at the five plaquettes, the basis \eqref{eq:basis} will give a flux-tube network different from the one in Fig. \ref{fig2}(b). However, this
ordering of the fluxons can be fixed once for all. All different choices of basis are
equivalent up to a unitary basis transformation, due to the topology symmetry of the states. (See the mutation symmetry in Ref.
\onlinecite{GSD}). The exclusion statistics can be conveniently analyzed in the above basis.
(The details are left to the forthcoming paper \onlinecite{Hu2}.)

The above analysis can be generalized to the case on a torus. The nontrivial topology
introduce two loops in the basis. The $N$-fluxon basis is expressed by \begin{equation}
\label{eq:TorusBasis} \TorusFluxonBasisA{\tau}{\tau}{\tau}{k_1}{k_2}{k_{N-1}}{p}
\quad\otimes\quad \TorusFluxonBasisB{\tau}{\tau}{\tau}{l_1}{l_2}{l_{N-1}}{q}. \end{equation}
Again, this can be viewed as the skeleton of the flux-tube network through plaquettes of a
torus graph, with left part corresponding to the layer outside the torus surface while the
right one inside the torus surface. If we imagine the embedding of the torus surface into the
$S^3$ manifold, we see that $S^3$ is cut along the surface into two solid torus, giving rise
to the two non-contractible loops in the above basis. We can apply similar analysis as in the
sphere case: now in formula \eqref{torusbinomcounting}, $N_1=0,1$ stands for the number of
$p=\mathbf{1}$ (and $N_2$ for the number of $q=\mathbf{1}$). The mutual exclusion
$\alpha_{31}=2=\alpha_{42}$ survives because of the exclusion between $p=0$ and $k_1=0$ (or
$k_{N-1}$=0).

In particular, the ground states have four-fold degeneracy, with the basis $\ket{p,q}$ as
special example of the formula \eqref{eq:TorusBasis}.


\begin{thebibliography}{100}
\bibitem{footnote} A different state-counting formula appeared in ref. \onlinecite{Vidal} for the particular case with Fibonacci data, but no connection to exclusion statistics was made.
\end{thebibliography}
\end{document}